\documentclass[aps,prl,twocolumn,groupedaddress,showpacs]{revtex4}

\usepackage{amsmath,amsthm,amsfonts,amscd} 
\usepackage{eucal}
\usepackage{graphicx}
\usepackage{color}		
\usepackage{ulem} 		
\usepackage{pxfonts}	
\usepackage{upgreek}	
\usepackage[usenames,dvipsnames]{xcolor}

\newcommand{\timout}[1]{\bgroup \markoverwith{\textcolor{magenta}-}\ULon{#1}}

\begin{document}

\title{Dynamics of the non-classical light from a single solid-state quantum emitter}

\author{Edward B. Flagg}
	\email{edward.flagg@nist.gov}
\author{Sergey V. Polyakov}
\author{Tim Thomay}
\author{Glenn S. Solomon}
	\email{glenn.solomon@nist.gov}
	\affiliation{Joint Quantum Institute, National Institute of Standards and Technology, 
				\& University of Maryland, Gaithersburg, MD, USA.}

\date{\today}

\begin{abstract} 
We measure the dynamics of a non-classical optical field using two-time second-order correlations in		
conjunction with pulsed excitation. The technique quantifies single-photon purity and coherence during		
the excitation-decay cycle of an emitter, illustrated here using a quantum dot.  We observe that for 		
certain pump wavelengths, photons detected early in the cycle have reduced single-photon purity and			
coherence compared to those detected later.  A model indicates that the single-photon purity dynamics 		
are due to exciton recapture after initial emission and within the same pulse cycle.						
\end{abstract}																					

	
\maketitle

Photons produced by spontaneous emission from a two-level quantum system have strongly sub-Poissonian statistics \cite{HBT1956Nat,Kimble77,Michler2000Sci} and are indistinguishable \cite{Hong1987PRL,Santori2002} if they are free of decoherence.  Such indistinguishable single-photon light is central to a variety of fundamental experiments \cite{Aspect82,Torgerson95,Pan00,Brida08} and emerging schemes in quantum information processing \cite{Ekert91,Knill01,Duan2001Nature,Duan04,Duan10}.  Single-photon purity \cite{HBT1956Nat} and coherence \cite{Hong1987PRL} are important properties of the light that are often treated statically in experiments.  Depending on the excitation and the environment, however, in many systems these properties will be dynamic.

Pulsed excitation of a quantum emitter results in non-classical light that is clearly time dependent.  Systems such as single atoms passing through optical cavities \cite{Haroche93}, strongly coupled cavity-emitter systems \cite{McKeever04,Reithmaier2004Nat}, and quantum memories capable of storing and retrieving single photon states \cite{Eisaman2005, Saglamyurek2011, Clausen2011}, can have complicated dynamics in the excitation and emission process. Therefore, characterizing the dynamics of the emitted light is salient to an improved understanding of both the emitted non-classical light field, and interactions between the emitter and environment.  

In typical second-order correlation measurements a stationary emission process is implied, and thus a single-time function, $g^{(2)}(\tau)$, is used, where $\tau$ is the difference between photon detection times on the two detectors \cite{Santori2002}.
For non-stationary processes, however, this obscures the dynamics of the individual fields.
Thus, a two-time correlation function, $g^{(2)}(t_{1},t_{2})$, should be used, where $t_{1}$ and $t_{2}$ are the delay times between the excitation pulse and photon detection on the two detectors. 
For example, such a correlation was measured previously on two already correlated thermal fields \cite{Polyakov2004PRL}.

Here we perform for the first time  $g^{(2)}(t_{1},t_{2})$ measurements on a single strongly sub-Poissonian light field.  
The measurements are performed on the emission from a single excitonic transition in a quantum dot (QD).  
We show instances where the time-averaged Hanbury Brown-Twiss (HBT) autocorrelation, $g^{(2)}_{\rm{HBT}}(\tau)$, would indicate that the QD decay produces single photon-like states \cite{Kimble77}, while time-dependent measurements, $g^{(2)}_{\rm{HBT}}(t_{1},t_{2})$, clearly show regions where it does not.  
We also determine the dynamical photon indistinguishability by interfering two replicas of the field in a Hong-Ou-Mandel (HOM) cross-correlation measurement \cite{Hong1987PRL}, denoted by $g^{(2)}_{\rm{HOM}}(t_{1},t_{2})$.  These measurements demonstrate how the state would perform in single-photon based quantum protocols and characterize newly revealed underlying physics in the emission of the studied emitters.

The sample was made using molecular-beam epitaxy and contains a low density (approximately 10 $\upmu$m$^{-2}$) of strain-induced InAs QDs. 
The QDs are located at an anti-node in a 4-$\lambda$ planar distributed Bragg reflector (DBR) microcavity with 15.5 lower (10 upper) DBR pairs of GaAs and AlAs; the cavity mode is centered at $\lambda = 920$ nm.  A single-mode optical fiber is bonded to the cleaved $[110]$ sample face to couple the excitation laser into the guided mode of the DBR cavity \cite{Muller2008PRL}.  The sample is maintained at 5 K in a cryostat and excited by a mode-locked Ti:sapphire laser with a repetition rate of 76.1 MHz (period $=$ 13.14 ns) and 8 ps pulse duration.  The QD emission peak we analyzed is at 917.5 nm and high-resolution spectral measurements with a Fabry-Perot interferometer show that it lacks any fine structure splitting, implying that it is due to charged exciton (trion) decay.  Several excitation wavelengths were used: above the GaAs band-gap at 755 nm, and two quasi-resonant excitations at 893.0 nm and 904.1 nm.  Quasi-resonant conditions excite quantum confined states that have non-radiative transitions to the QD trion.

\begin{figure}
	\includegraphics[width=3.38in,angle=0]{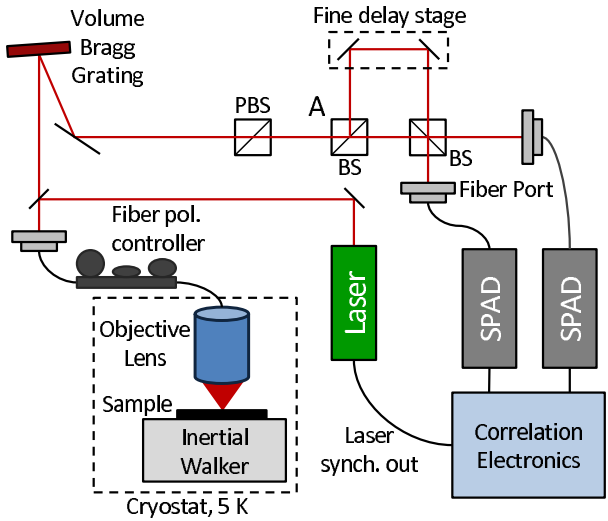}
 	\caption{Schematic of the experimental setup. 
 	The QD microcavity is kept at 5 K.  Its emission is fiber coupled to second-order auto-correlation measurements: With beamsplitter 'A' removed the geometry is suitable for measuring photon statistics; with 'A' in place an asymmetric Mach-Zender interferometer is created and the emissions from different excitation pulses meet at the beam splitter for second-order cross-correlation measurements of the photon indistinguishability.  In both cases detections at the beam splitter outputs are made with single-photon avalanche detectors (SPADs).
	}
	\label{f:Schematic}
\end{figure}

The emission from the QD is collected by a fiber-coupled objective lens and sent to the correlation setup shown in Fig.~\ref{f:Schematic}.
To measure the time-dependence of the multi-photon emission probability, we remove the first beamsplitter of the interferometer, marked 'A' in Fig.~\ref{f:Schematic}, to obtain an HBT configuration.
To interfere two replicas of the field in a HOM cross-correlation measurement the beam splitter 'A' is inserted to establish an asymmetric Mach-Zender interferometer.  One arm of the interferometer is 3 pulse periods (39.4 ns) longer than the other so the emissions from two different excitation pulses meet simultaneously at the final beamsplitter.
The outputs of the beamsplitter are each sent to a single-photon avalanche detector (SPAD) with a time response full-width half-maximum of 560 ps, that record time differences  $t_{1}$ and $t_{2}$ between each detection and a synchronous signal from the mode-locked laser.  From this raw data we construct the two-time second-order correlation of the light, $g^{(2)}(t_{1},t_{2})$.

\begin{figure}
	\includegraphics[width=3.38in,angle=0]{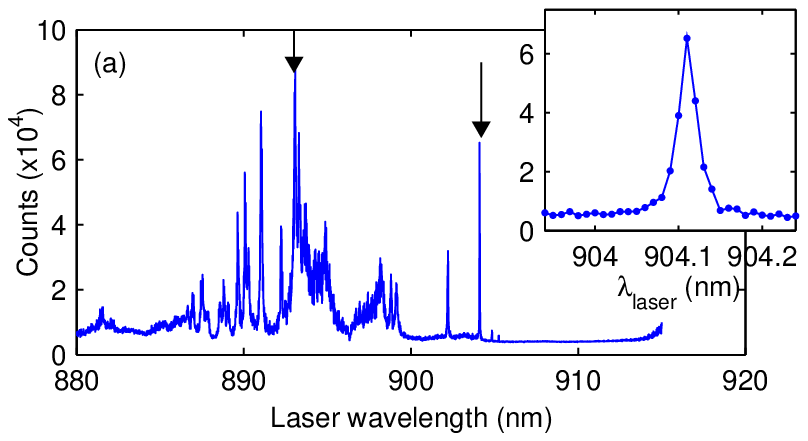}
	\includegraphics[width=3.38in,angle=0]{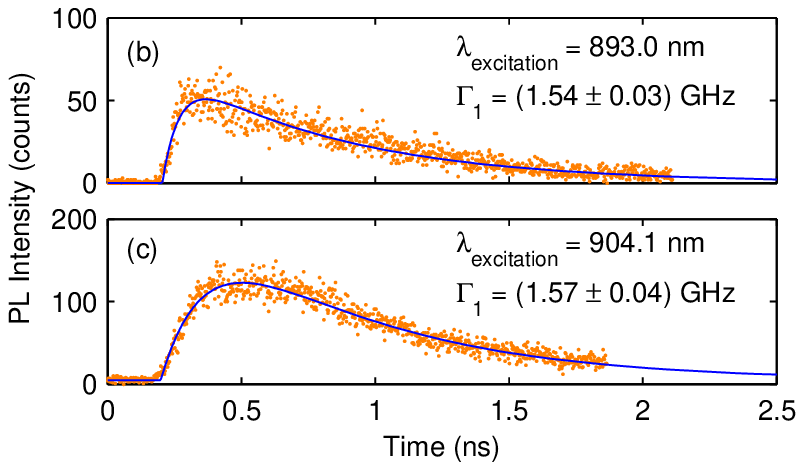}
	\caption{(Color online) (a) Photoluminescence excitation (PLE) spectrum of a QD trion peak at 917.5 nm.  Arrows indicate the quasi-resonant excitation wavelengths 893.0 nm and 904.1 nm used in the time-dependent measurements in (b) and (c).  The insert shows an expanded view of the PLE peak at 904.1 nm. (b)-(c) Time-dependent emission from the trion decay where the excitation wavelength is indicated.  The lifetimes are fit with exponential decay.  The characteristic rise times in trion population are (14.4$\pm$1.5) GHz (893.0 nm) and (4.4$\pm$0.3) GHz (904.1 nm).}
	\label{f:PLE}
\end{figure}

Figure \ref{f:PLE}(a) shows a photoluminescence excitation (PLE) spectrum of the QD emission peak of interest; the inset shows an expanded view of the PLE peak at 904.1 nm.  Figures \ref{f:PLE}(b) and (c) show lifetime measurements recorded with a streak camera with 30 ps time resolution for quasi-resonant excitation at 893.0 nm and 904.1 nm respectively, indicated in Fig.~\ref{f:PLE}(a) by arrows.   
The data are fit by functions of the form $[1-\rm e^{-\Gamma_{\rm{rise}} t}] \rm e^{-\Gamma_{1} t}$; the decay rates, $\Gamma_1$, for the two cases are equal to within the experimental error; $\Gamma_{1} = (1.56 \pm 0.05)$ GHz. The rise rates, $\Gamma_{\rm{rise}}$, are $(14.4 \pm 1.5)$ GHz for 893.0 nm excitation, and $(4.4 \pm 0.3)$ GHz for 904.1 nm excitation.
The rise rate is faster for 893.0 nm because at this wavelength the laser excites a sharp PLE line plus additional states of the wetting layer tail, and therefore the exciton state can be populated through several channels.
Conversely, there is likely only a single state associated with the the 904.1 nm excitation.

\begin{figure*}[t]
	\includegraphics[height=3.5in,angle=0]{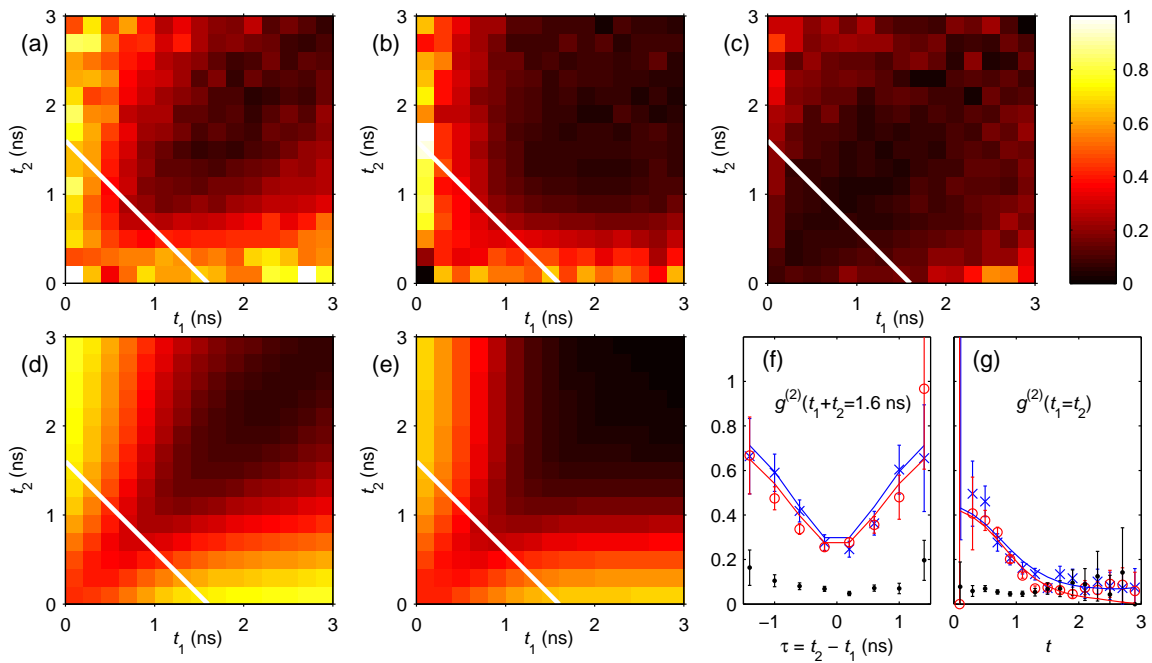}
	\caption{(Color online) Two-time second-order auto-correlations $g^{(2)}_{\rm{HBT}}(t_{1},t_{2})$. 
		Experimentally observed correlations for (a) 755 nm, (b) 893.0 nm, and (c) 904.1 nm excitation, respectively.
		Simulations of the correlations for (d) 755 nm and (e) 893.0 nm.
		Cuts of the $g^{(2)}_{\rm{HBT}}(t_{1},t_{2})$ surface (f) along the solid white lines, $g^{(2)}_{\rm{HBT}}(t_{1}+t_{2}=1.6)$, and (g) along the diagonal, $g^{(2)}_{\rm{HBT}}(t_1=t_2)$ for excitation at 755 nm (blue crosses), 893.0 nm (red circles) and 904.1 nm (black dots).
		Errorbars indicate the 95\% confidence range, and lines are the simulations.
	}
	\label{f:Autocorr}
\end{figure*}

Figures~\ref{f:Autocorr}(a)-(c) show $g^{(2)}_{\rm{HBT}}(t_{1},t_{2})$ for the three excitation wavelengths 755 nm, 893.0 nm, and 904.1 nm, respectively. For excitations at 755.0 and 893.0 nm the value of $g^{(2)}_{\rm{HBT}}(t_{1},t_{2})$ is largest when either $t_{1} \approx 0$, $t_{2} \approx 0$ or $t_{1},t_{2}\approx 0$, {\it i.e.}~when at least one photon is detected early in the exciton lifetime.  
This indicates that after the excitation pulse, carriers remain that can be captured in the QD allowing subsequent emission of another photon. 
In both Figs.~\ref{f:Autocorr}(a) and (b), for small values of $t_1$ and $t_2$, $g^{(2)}_{\rm{HBT}}(t_{1},t_{2}) > 0.5$, which is not below the threshold for single-photon purity \cite{Kimble77}.
The time-averaged values, however, are 0.31, 0.18, and 0.11 for Figs.~\ref{f:Autocorr}(a), (b), and (c), respectively.  Comparison of the time-averaged and time-dependent $g^{(2)}_{\rm{HBT}}$ values demonstrates that a single-photon like time-averaged $g^{(2)}_{\rm{HBT}}$ does not guarantee that the emitter acts like a single-photon emitter over the entire decay.
Also note the differences between these cases: the shape of the region with lowest value in Fig.~\ref{f:Autocorr}(a) resembles an oval whose long axis is the $t_1=t_2$ line, while in Fig.~\ref{f:Autocorr}(b) this region appears rectangular. This is due to different scenarios for additional exciton capture. 
In contrast to both 755 nm and 893.0 nm, excitation at 904.1 nm (Fig.~\ref{f:Autocorr}(c)), results in a relatively flat and low $g^{(2)}_{\rm{HBT}}(t_{1},t_{2})$ because there is only one excitation channel for the exciton and thus little possibility to capture another excitation within the pumping cycle.


Much of the behavior of $g^{(2)}_{\rm{HBT}}(t_{1},t_{2})$ in Fig. 3 is related to the occupation and decay dynamics of various excited carrier reservoirs in the GaAs and InGaAs wetting layer, which are populated differently depending on the excitation wavelength.
Over time these carriers either are captured in the QD or decay by other channels such as recombination through bulk exciton states or wetting layer states.  If a photon is emitted before the carrier reservoir populations have decayed, then a second exciton can be captured by the QD and can be emitted at a later time.

A model to describe this process is developed and discussed in the Supplemental Material.  The simulations in Fig. 3 are based on this model.
The model shows that in the limit when the capture rate is large compared to the emission rate, the source is in general not anti-bunched.  It approaches Poissonian except in a valley along $t_1=t_2$ of width characterized by the average capture time.  In the center of the valley $g^{(2)}_{\rm{HBT}}(t_{1},t_{2})$ ideally goes to zero but will in practice be higher due to convolution with the detector response.  In contrast, when the capture rate is small compared to the emission rate, the source becomes anti-bunched.  In Fig. 3(d)-(e) at short time periods, the capture rates are similar to the emission rate, therefore the values of $g^{(2)}_{\rm{HBT}}(t_{1},t_{2})$ are large.  At long time periods, the capture rate is much less than the emission rate, therefore $g^{(2)}_{\rm{HBT}}(t_{1},t_{2})$ approaches zero.

Figure~\ref{f:Autocorr}(f) is a cut of the $g^{(2)}_{\rm{HBT}}(t_1,t_2)$ surface along $t_2=t_{\rm{const}}-t_1$ for each excitation wavelength.  For 755 nm and 893.0 pump wavelengths, the cut shows the anti-bunching valley at $t_1=t_2$.  The depth and width of the valley increases as $t_{\rm{const}}$ increases.  For the 904.1 nm pump wavelength, the value is low over the whole range.
In previous work where only a one-dimensional $g^{(2)}_{\rm{HBT}}(\tau)$ is considered for a pulsed source, the dynamics, especially the central valley, are lost due to time averaging.  These pump-dependent features are clearly identifiable in our experiment, although some fast dynamics are obscured by the jitter of the detectors.
Figure~\ref{f:Autocorr}(g) is a cut along $t_1=t_2$ which shows the time dependence of $g^{(2)}_{\rm{HBT}}(\tau=0)$ after the pump pulse.  Though the model predicts the value of $g^{(2)}_{\rm{HBT}}(\tau=0)$ is zero, convolution with the detectors' time response results in a higher value.
For both 755 nm and 893.0 nm excitation wavelengths the value decreases over time owing to a decay of carriers in the reservoirs. In a typical single-time $g^{(2)}_{\rm{HBT}}(\tau)$ measurement, this dependence is irrecoverable. For quasi-resonant 904.1 nm excitation, to within the measurement uncertainty $g^{(2)}_{\rm{HBT}}(\tau=0)$ does not depend on time.

The $g^{(2)}_{\rm{HBT}}(t_1,t_2)$ data for 755 nm and 893.0 nm excitation in Figs.~\ref{f:Autocorr}(a) and (b) can be fit using the above-mentioned model and parameters extracted from the measurements in Fig.~\ref{f:PLE}.  The simulated $g^{(2)}_{\rm{HBT}}(t_1,t_2)$ are presented in Figs.~\ref{f:Autocorr}(d) and (e) where the effects of detector jitter are included.  
The values of the fit parameters are detailed in the Supplemental Material.  The fit for 893.0 nm excitation requires one carrier reservoir, but the explanation of the 755 nm excitation is more complex, requiring multiple reservoirs.  For 755 nm excitation we observe a fast initial decay of $g^{(2)}_{\rm{HBT}}(t_1,t_2)$, consistent with a short lifetime of free carriers, but instead of the rectangular valley seen with 893.0 nm excitation an oval region of the lowest $g^{(2)}_{\rm{HBT}}(t_1,t_2)$ values occurs. This is consistent with a small number of long-living carriers, possibly due to shallow traps in the vicinity of the QD. 
The relative contribution of long-lived carriers becomes significant at later times, contributing to a visible change in shape of the observed $g^{(2)}_{\rm{HBT}}(t_1,t_2)$.  Although the population of long-lived carriers is two orders of magnitude lower than that of the short-lived carriers, both contributions are resolvable.  Thus, even weak effects of the environment on a quantum emitter can be readily detected by a two-time resolved measurement.
Peter et al.~\cite{Peter2007APL} reported that emitters with short lifetimes that are pumped through an incoherent capture process have significantly larger multi-photon emission probability overall.  Here we experimentally show that an incoherently pumped emitter with \textit{any} lifetime has a significant probability for multi-photon emission at the beginning of the emission cycle, as its time-resolved $g^{(2)}_{\rm{HBT}}$ changes from unity to nearly zero in every excitation cycle.

We characterized the coherence properties of the field emitted by the QD using HOM cross-correlation.  If one photon is incident on the beamsplitter from each of the two inputs, and these photons are indistinguishable, then there will be a two-photon Fock state at one output and no photons at the other.  This bunching process has been called coalescence \cite{Flagg2010PRL,Bylander2003,Legero2004PRL}.  
An expression for the coalescence can be derived from the definition of the HOM cross-correlation for the special case of identical emitters: $ C(t_{1},t_{2}) = 1+g^{(2)}_{\rm{HBT}}(t_{1},t_{2}) - 2 g^{(2)}_{\rm{HOM}}(t_{1},t_{2})$, where the subscripts indicate the configuration of the measurement.  
The coalescence measures the degree of indistinguishability of the emission from separate excitations, conditional on detecting one photon at time $t_{1}$ and another at time $t_{2}$ after the laser pulse.  Because the expression includes $g^{(2)}_{\rm{HBT}}(t_{1},t_{2})$, it eliminates the statistical effects of multi-photon emission.  Therefore the coalescence can be viewed as a measure of the indistinguishability of emission from consecutive excitations, regardless of the photon statistics, and $C(t_{1},t_{2})$ has a range $0\leq C(t_{1},t_{2}) \leq 1$: 0 for distinguishable photons, and 1 for indistinguishable ones. 
 
\begin{figure}[t]
	\includegraphics[width=3.38in,angle=0]{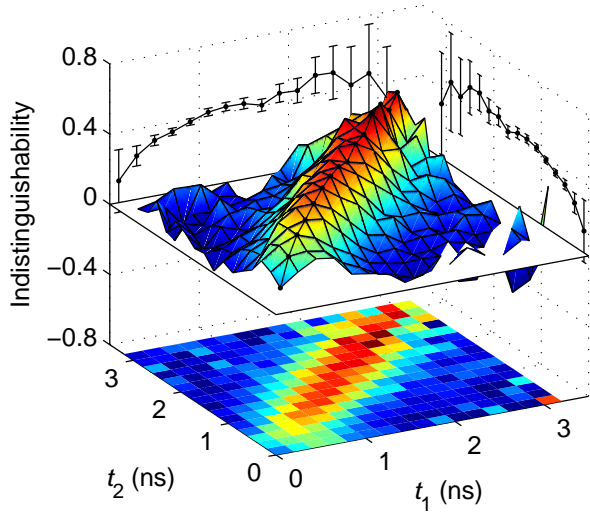}
	\caption{Single QD indistinguishability measured through the coalescence probability, $C(t_{1},t_{2})$. The pulsed excitation is at 755 nm.}
	\label{f:Indist}
\end{figure}

Figure~\ref{f:Indist} shows $C(t_{1},t_{2})$ for excitation at 755 nm. 
As expected for a state with decoherence, the highest coalescence probability is a ridge along $t_{1} = t_{2}$, where $C(t_{1} = t_{2})=1$ for infinitely fast detectors.  $C(t_1,t_2)$ decreases monotonically away from $t_{1} = t_{2}$, with the width of the ridge proportional to the photons' coherence time.
With our detectors, $C(t_1,t_2)$ gets averaged over the jitter time of the detectors and the maximum value becomes proportional to the width of the ridge. The coalescence values along $t_{1} = t_{2}$ are projected onto the rear planes of the figure, and we see that $C(t_{1},t_{2})$ is smaller for small $t_{1}$ and $t_{2}$ then for later times.  Therefore during the transient time after the initial pump pulse the indistinguishability of the state is significantly reduced. 
This behavior results from variations in the wavefunction overlap of the two photons.  We attribute it to the variation, or jitter, in exciton capture time due to the incoherent excitation.  The reduction could also be due to extra dephasing from non-equilibrium carrier population around the QD.

The experimental result above demonstrates the average coherence loss associated with incoherent excitation first predicted by Kiraz et al.~\cite{Kiraz2004PRA}.
It was pointed out that for high QD decay rates, indistinguishability will be degraded when using incoherent pumping because the jitter in the capture time becomes comparable to the lifetime.  Here we observe an additional dynamical effect for short measurement times.

Using a new experimental technique we characterize the non-classical dynamics of a single quantum emitter. 
We show that both auto-correlation and cross-correlation data have temporal structure which depends on the pumping method used. 
In several pumping schemes, the dynamics of auto-correlation are consistent with a multi-carrier capture model. 
We show situations where typical time-difference measures indicate a single-photon character in the emission statistics while the full dynamics indicate temporal regions where it is not.
This result not only bears on optically excited emitters, but also electrically excited single emitters because such processes are based on probabilistic capture from a reservoir.
The dynamics of the cross correlation is also affected by the pulsed incoherent pumping, where a temporal dependence in the indistinguishability is observed.

We acknowledge partial support from the NSF Physics Frontier Center at the Joint Quantum Institute.  T. Thomay acknowledges support through the NIST-ARRA Fellowship program at UMD.

\end{document}